\documentclass{cimento}[]

\usepackage{graphicx}  

\title{Non-strange light-meson spectroscopy at COMPASS}
\author{
    P.~Haas on behalf of the COMPASS collaboration
    \thanks{philipp.haas@tum.de}
}
\instlist{
    \inst Phys. Dept. E18, TU M\"unchen, Garching, Germany
}
\begin{document}
\maketitle

\begin{abstract}
Lattice-QCD predicts the exotic meson $\pi_1(1600)$ to dominantly decay to $b_1\pi$. The $b_1\pi$ decay channel is accessible via the $\omega\pi^{-}\pi^{0}$ final state. COMPASS recorded the so far largest data set of this final state. A partial-wave analysis allows to determine the resonant content in this final state including possible contributions from $\pi_1(1600)$. Decomposing the measured intensity into amplitudes of partial waves gives a first qualitative insight into contributing intermediate states. We observe signals in agreement with well-established states like the $\pi(1800)$ and $a_4(1970)$. Smaller resonance-like signals are visible in the $J^{PC}$ sectors $3^{++}$ and $6^{++}$, where possible states were claimed but none are established. For $J^{PC}=1^{-+}$ a signal at $1.65\,\mathrm{GeV/}c^{2}$ in $b_1(1235)\pi$ partial waves is consistent with the expected $\pi_1(1600)$.
\end{abstract}

\section{Introduction}
The constituent quark model describes mesons as $q\bar{q}$ bound states, systematically following a multiplet structure derived from basic symmetries. However, QCD allows further states beyond this $q\bar{q}$ configuration. Other possible states --- so-called exotic mesons --- are hybrids, glueballs, and multiquark-states. Mesons with $J^{PC}$ quantum numbers forbidden for a conventional $q\bar{q}$ state, like $J^{PC}=1^{-+}$, are called spin-exotic mesons. Lattice-QCD  predicts the lightest hybrid state as a single pole with $J^{PC}=1^{-+}$~\cite{Dudek:2013yja}. Thanks to recent advances, lattice-QCD also predicts the partial decay widths of this pole from first principle~\cite{Woss:2020ayi}, where $b_1\pi$ is the dominant decay channel. Other channels like $\rho\pi$, $f_1(1285)\pi$, $\eta^{(}\vphantom{}'^{)}\pi$, and $K^*\bar{K}$ should be suppressed by about an order of magnitude and the final state $\rho\omega$ is predicted to contribute less than $1\%$ to the total decay width.

Experimentally, $\pi_1$ signals were observed in different decay modes at masses of \linebreak$1.4\,\mathrm{GeV}/c^2$ and $1.6\,\mathrm{GeV}/c^2$. As a result, two $\pi_1$ states were claimed, namely $\pi_1(1400)$ and $\pi_1(1600)$. A coupled-channel analysis of $\eta\pi$ and $\eta'\pi$ using COMPASS data demonstrated that a single pole is sufficient to describe the partial waves in both decay channels~\cite{Rodas:2018owy}. COMPASS further observed the $\pi_1(1600)$ decaying to $\rho\pi$ and $f_2(1270)\pi$~\cite{Akhunzyanov:2018lqa}.

Here, we present a new study of the $b_1\pi$ decay mode of $\pi_1(1600)$ at COMPASS which requires a partial-wave analysis of the $\omega\pi^{-}\pi^{0}$ final state. We present the recent status of this analysis.

\section{Analysis of the $\omega\pi\pi$ final state}
\label{sec:analysis}

At COMPASS excited mesons $X^{-}$ are produced by diffractive scattering of an $190\,\mathrm{GeV}$ $\pi^{-}$ beam off a liquid-hydrogen target. At such energies, Pomeron exchange is expected to be the dominant production mechanism. This process gives access to excited intermediate states $X=a_J,\pi_J$.  As states in the light-meson sector overlap, a partial-wave analysis is necessary to disentangle the different contributing $X^{-}$ mesons. It also allows to measure the quantum numbers of the states. We model the intermediate states --- the partial waves --- using the isobar model, where $X^{-}$ decays to $\omega\pi\pi$ in successive two-body decays. To uniquely identify a particular partial wave, a set of quantum numbers
\[i=J^PM^\epsilon[\xi l]b LS\]
is necessary. Here, $J$ is the total spin of $X^{-}$, $P$ is the parity, and $M^{\epsilon}$ characterises the spin projection of $X^{-}$ on the beam axis. $L$ and $S$ are the orbital angular momentum and intrinsic spin of the $X\to\xi b$ decay, respectively. $\xi$ is the so-called isobar, an intermediate state in a two-body subsystem of $\omega\pi\pi$, which is modelled using known resonances. For $\omega\pi^{-}\pi^{0}$, the possible two-body subsystems are $\pi^{-}\pi^{0}$, $\omega\pi^{-}$, or $\omega\pi^{0}$. We consider the isobars $\rho(770)$, $\rho(1450)$, and $\rho_3(1690)$ for the $\pi^{-}\pi^{0}$ intermediate state and $b_1(1235)$, $\rho(1450)$, and $\rho_3(1690)$ for the $\omega\pi$ intermediate states. Further, $l$ is the orbital angular momentum between the two daughters of the isobar. $b$ is the bachelor particle, e.g.\ the remaining particle in $\omega\pi\pi$ outside of the isobar. Two partial waves that differ only in the charge of $\xi$ and $b$, i.e.\ $\xi b = b_1(1235)^-\pi^0$ and $\xi b = b_1(1235)^0\pi^-$, are expected to have the same amplitude and are combined into one partial wave.

The description of the final state of $X^-$ requires a set of 8 phase-space variables $\tau$. The description of the decay $X^-\to\omega\pi^-\pi^0$ via an intermediate state $\xi$ requires the mass $m_\xi$ of the intermediate state and two two-particle decays, each described by two angles $\phi$ and $\theta$. In addition the decay of $\omega\to\pi^-\pi^0\pi^+$ requires a mass $m_\omega$ of $\omega$ and two Dalitz-plot variables.

Following the method presented in ref.~\cite{Akhunzyanov:2018lqa}, we model our measured intensity $\mathcal{I}$ as
\begin{equation}
    \mathcal{I}(m_{X},t',\tau)=\bigg| \sum_i \mathcal{T}_i(m_X,t') \psi_i(m_X,\tau) \bigg|^2.
    \label{eq:intensityModel}
\end{equation}
The intensity depends on the invariant mass $m_{X}$ of the excited $X^{-}$ state, the squared four-momentum transfer $t'$, and the phase-space variables $\tau$. In eq.~\ref{eq:intensityModel} we calculate the decay amplitude $\psi_i(m_X,\tau)$ using the isobar model. We model the mass line-shapes of isobars as Breit-Wigner resonances with dynamic width using the parameters from ref.~\cite{Workman:2022ynf}. The transition amplitude $\mathcal{T}_i$ contains all information about the production, the propagation, and the coupling of $X^{-}$ to a partial-wave $i$. We obtain its complex value by fitting eq.~\ref{eq:intensityModel} to the measured intensity distribution in a maximum-likelihood fit. We fit $\mathcal{T}_i$ as independent constants in cells of $m_X$ and $t'$. With this approach, we approximate $\mathcal{T}_i(m_X,t')$ as step-wise constant function without prior knowledge about any resonant content of $X^-$ in $i$.

To perform the fit, one must choose a finite subset of the infinite number of partial waves $i$ used in eq.~\ref{eq:intensityModel}. Traditionally, this has been done by manually selecting waves based on the expected strength of $\mathcal{T}_i$. To reduce potential bias in this selection, we developed an alternative approach at COMPASS based on regularisation-based model-selection techniques. In this approach, a large wave pool is constructed based on loose systematic constraints. In our case, we consider all waves with $J\leq8$, $M\leq2$, $L\leq8$, and $\epsilon=+1$ using all aforementioned isobars. This results in a total of 893 partial-waves. We fit all considered waves for each $(m_X,t')$ cell using a regularised likelihood. Due to the regularisation most of the waves are close to zero. The remaining waves are used as the wave set that is unique for each cell in $(m_X,t')$. By refitting the data with the selected waves and no regularisation we decompose the measured intensity into the amplitudes $\mathcal{T}_i$ of all significant partial waves $i$. 

\subsection{Results}
\label{sec:results}

To discuss the results, we focus on the intensity $\big|\mathcal{T}_i\big|^2$ and the phase $\arg \mathcal{T}_i$ of a partial wave as a function of $m_X$. Since the total phases are not measurable, we consider the phase difference between two partial waves $\phi=\arg (\mathcal{T}_i-\mathcal{T}_j)$. For an isolated pole far from thresholds one expects a Breit-Wigner resonance characterised by a peak in the intensity and a phase motion of $180^{\circ}$ around the resonances mass. We observe clear signals for the established states $\pi(1800)$ and $a_4(1970)$. In the $2^{++}$ and $2^{-+}$ sectors the picture is more complex as multiple established states overlap.

\begin{figure}
    \centering
    \includegraphics[height=0.4\textwidth]{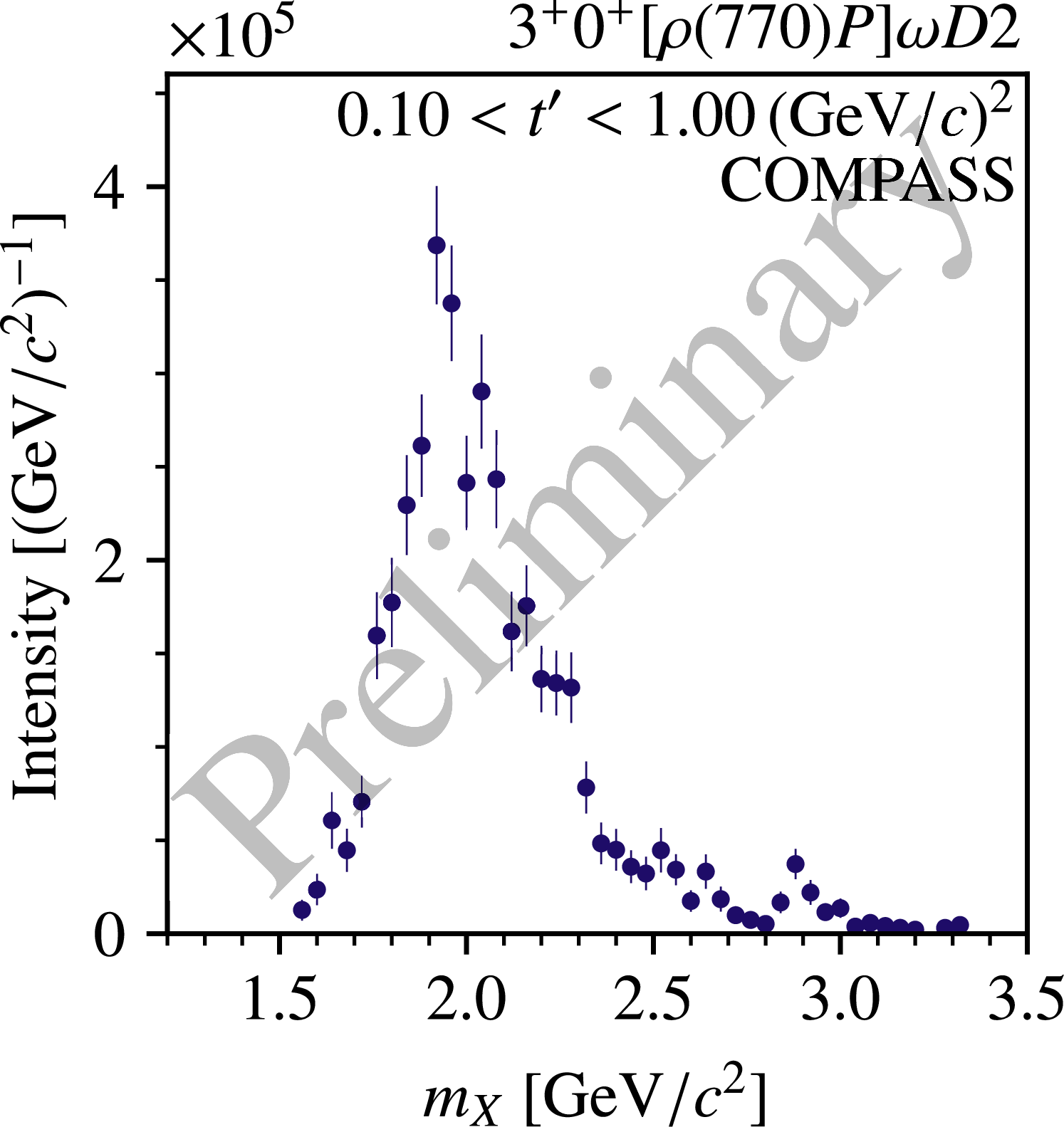}
    \includegraphics[height=0.4\textwidth]{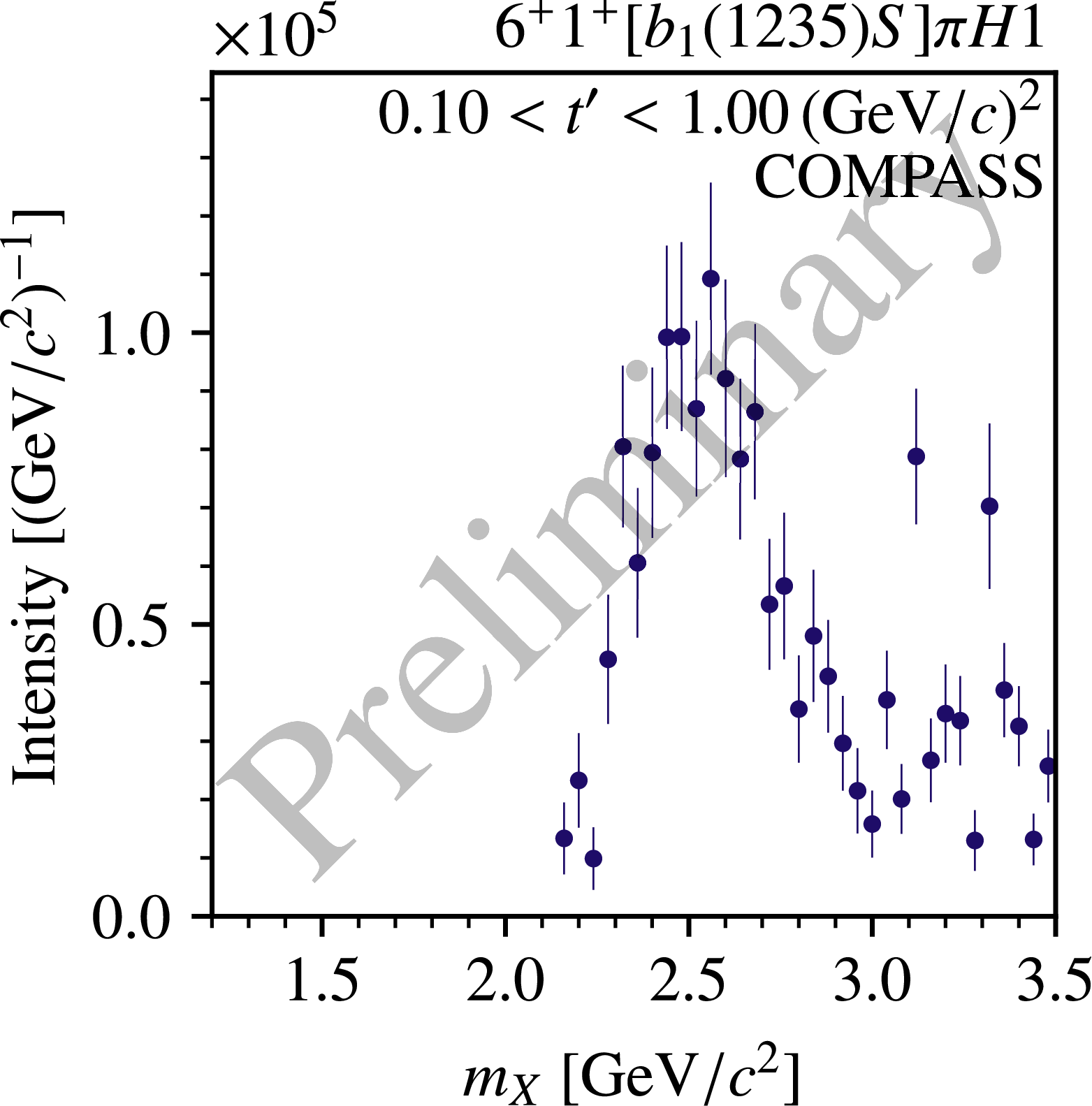}
    \caption{Intensities of the $3^+0^+[\rho(770)P]\omega D2$ (\textit{left}) and $6^+0^+[b_1(1235)S]\pi H1$ (\textit{right}) partial waves.}
    \label{fig:3++_6++}
\end{figure}

In the intensity of the $3^+0^+[\rho(770)P]\omega D2$ wave, a clear peak arises around $m_X=2.0\,\mathrm{GeV}/c^2$, shown in fig.~\ref{fig:3++_6++}~(\textit{left}). Consistent signals are observed in the decays into $b_1\pi$ and $\rho_3(1690)\pi$. The PDG lists two unconfirmed states at nearby masses, the $a_3(1874)$ and $a_3(2030)$. This would be the first observation of an $a_3$ in these decay channels.

Another sector in which we observe a resonance-like signal is the $6^{++}$ sector. Figure~\ref{fig:3++_6++}~(\textit{right}) shows the intensity of the $6^+1^+[b_1(1235)S]\pi H1$ wave. We observe a clear peak around $m_{X}=2.5\,\mathrm{GeV}/c^2$, consistent with the $a_6(2450)$ listed in the PDG. This state has only been seen in one experiment in $K_S K$~\cite{Cleland:1982te} and requires further confirmation. This would be the first observation in the $b_1(1235)\pi$ decay channel.

\begin{figure}
    \centering
    \includegraphics[height=0.4\textwidth]{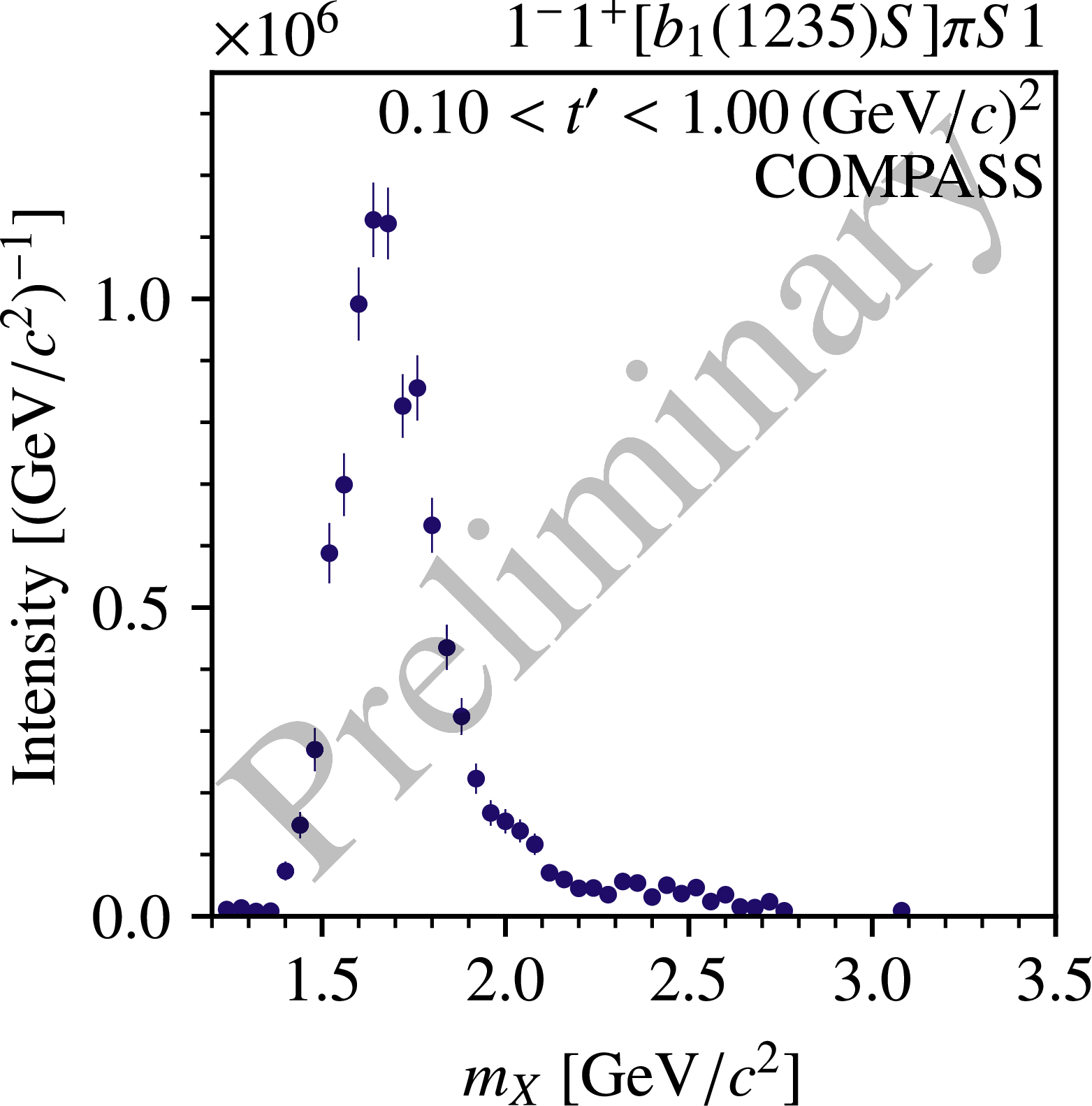}
    \includegraphics[height=0.4\textwidth]{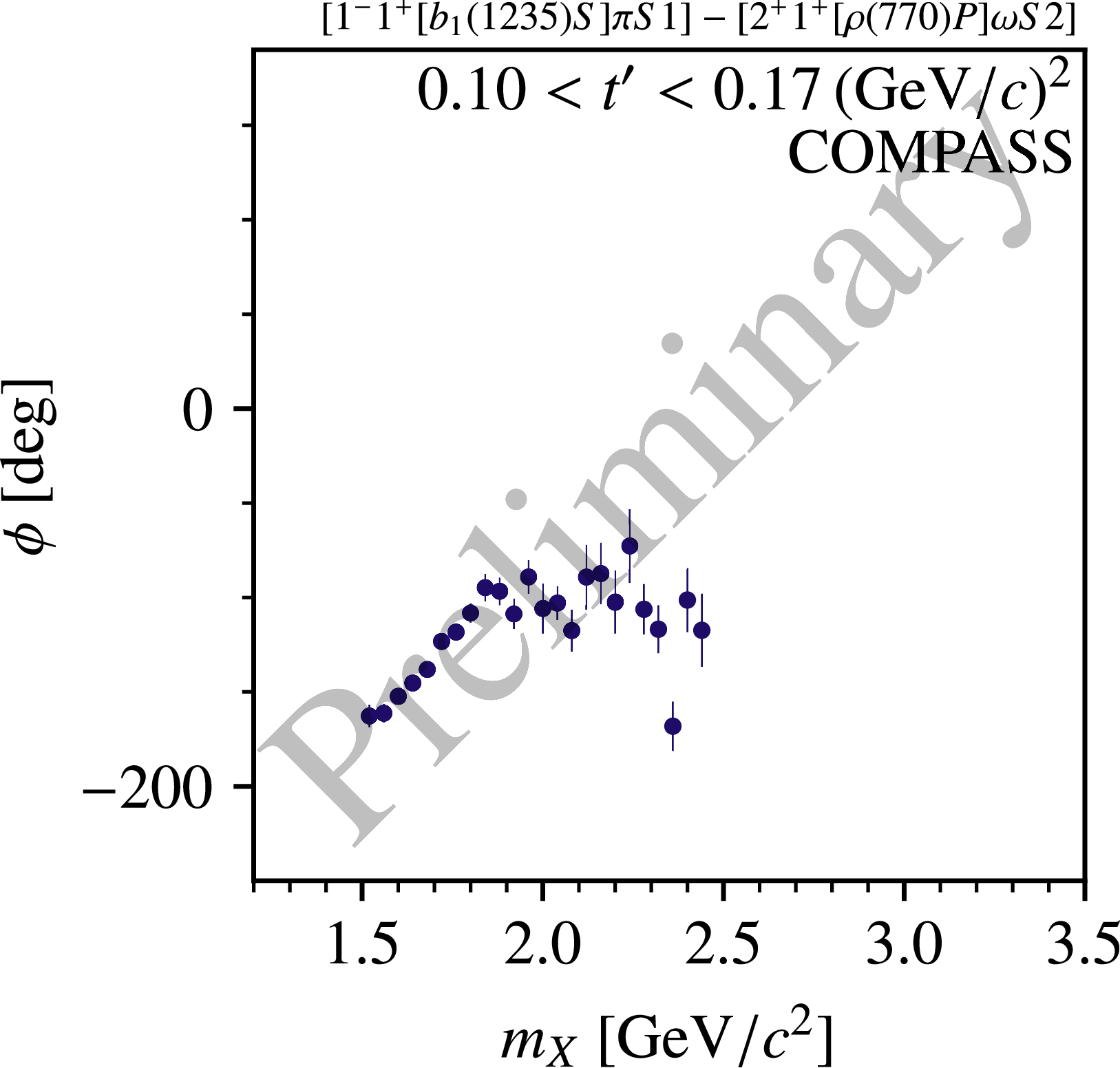}
    \caption{ \textit{Left:} Intensity of the $1^-1^+[b_1(1235)S] \pi S1$ partial wave. \textit{Right:} phase of the same wave w.r.t.\ the $2^+1^+[\rho(770)P]\omega S2$ wave.}
    \label{fig:1-+_b1PiS}
\end{figure}

\begin{figure}
    \centering
    \includegraphics[height=0.4\textwidth]{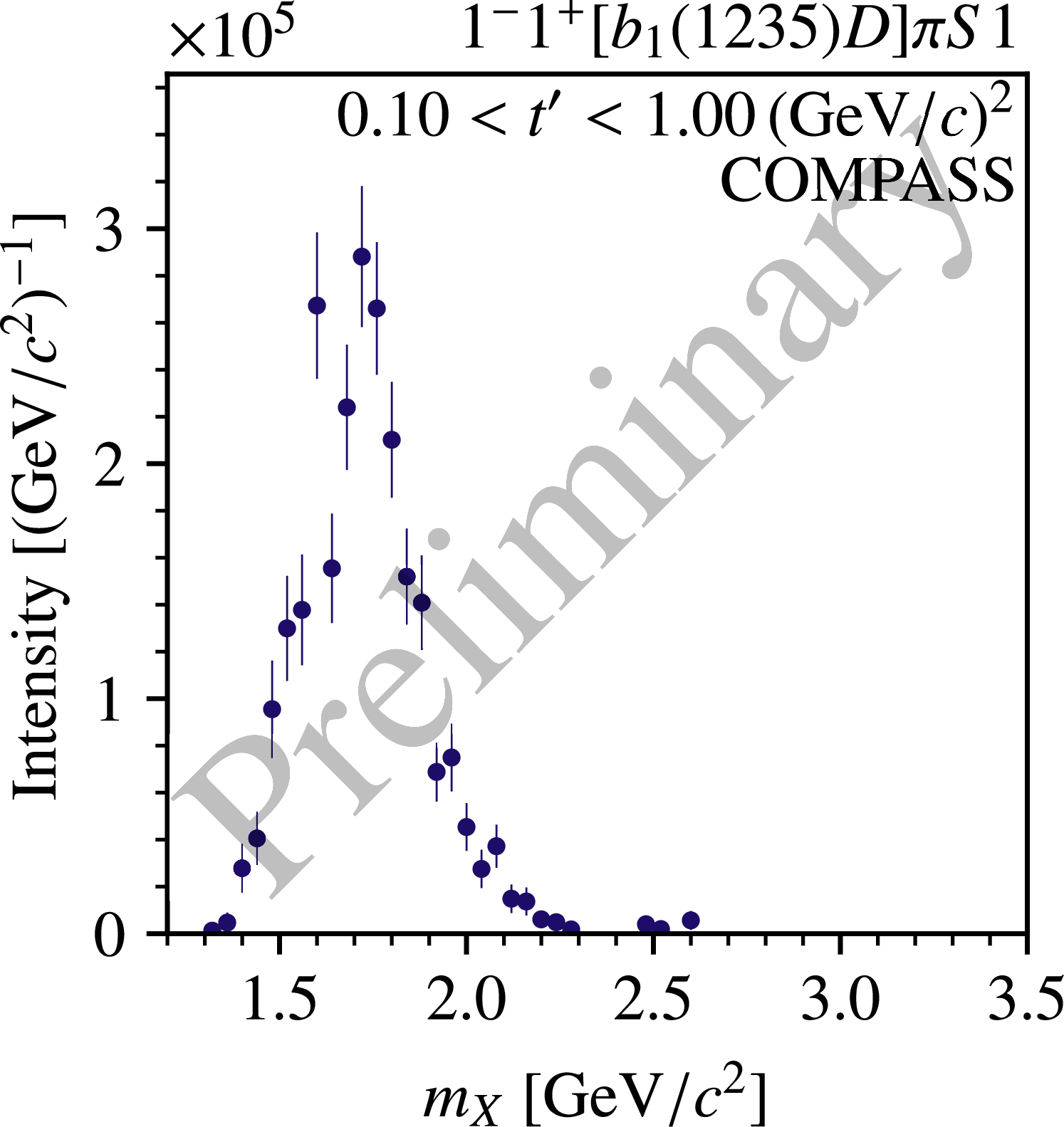}
    \includegraphics[height=0.4\textwidth]{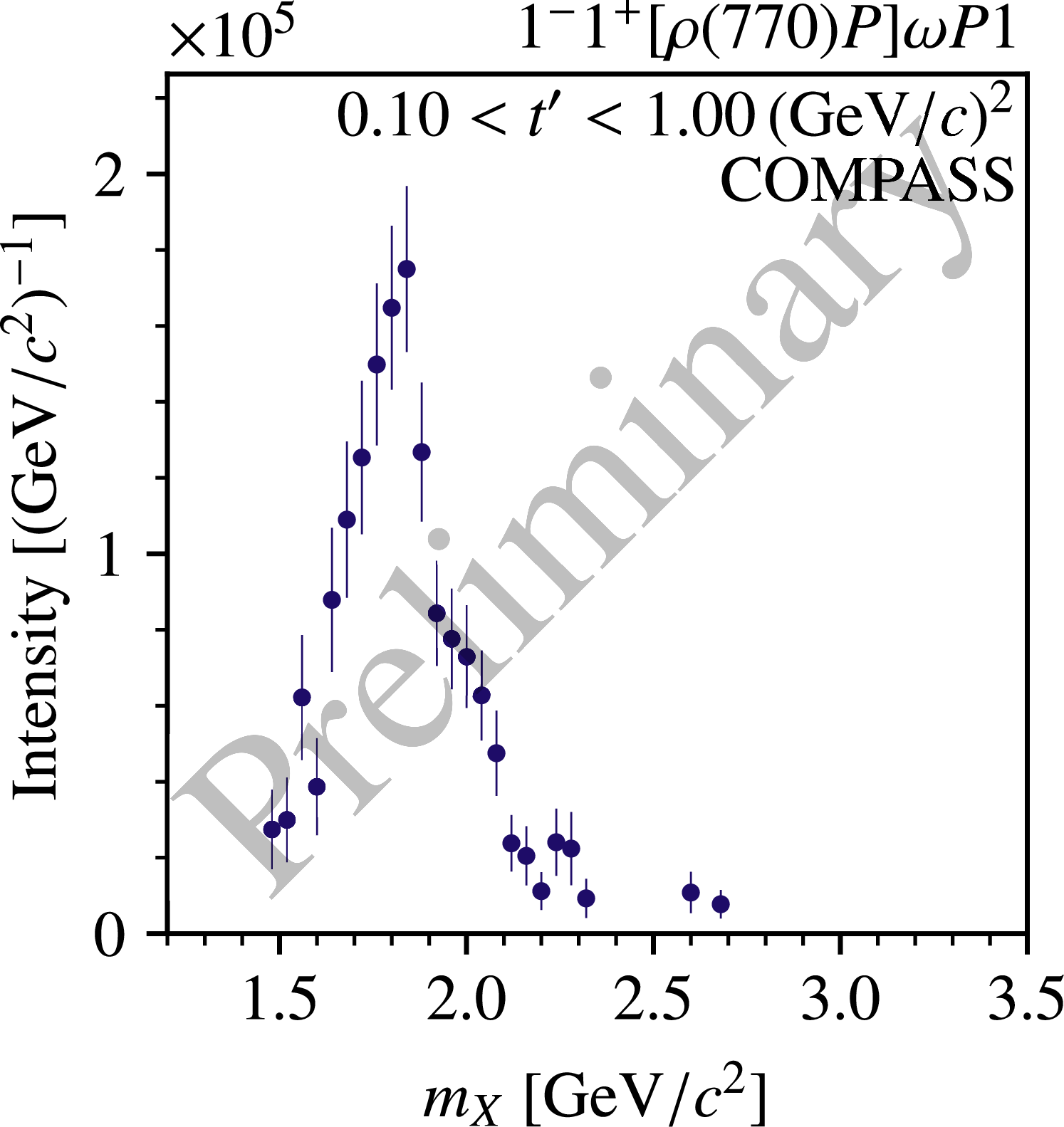}
    \caption{Intensities of the $1^-1^+[b_1(1235)D] \pi S1$ (\textit{left}) and $1^-1^+[\rho(770)P]\omega P1$ (\textit{right}) partial waves.}
    \label{fig:1-+_addWaves}
\end{figure}

In the $1^{-+}$ sector, we expect to observe a strong contribution from the $\pi_1(1600)$ in $b_1\pi$ waves. Figure~\ref{fig:1-+_b1PiS} shows the intensity and phase of the $1^-1^+[b_1(1235)S] \pi S1$ wave. A signal close to $m_{X}=1.6\,\mathrm{GeV}/c^2$ suggests a contribution from the well-established spin-exotic meson. We observe a similar signal in the wave where $b_1(1235)$ decays not via $S$- but $D$-wave, as shown in fig.~\ref{fig:1-+_addWaves}~(\textit{left}). This signal is consistent with the $\pi_1(1600)$ observed at COMPASS in $\rho\pi$ and $\eta'\pi$ decays. We also observe a resonance-like signal in $1^{-+}$ $\rho\omega$-waves, which is shown in fig.~\ref{fig:1-+_addWaves}~(\textit{right}). The $1^-1^+[\rho(770)P] \omega P1$ wave exhibits a signal around $m_{X}=1.8\,\mathrm{GeV}/c^2$. This signal could correspond to the $\pi_1(1600)$ with a shifted mass due to limited phase space. It would be the first observation of the $\pi_1(1600)$ in $\rho(770)\omega$ and would be inconsistent with the prediction of a small $\rho\omega$ partial decay width of $\pi_1(1600)$.

\section{Conclusion and outlook}
\label{sec:conclusion}

We have decomposed the COMPASS data selecting the $\omega\pi^{-}\pi^{0}$ and observe clear signals in several partial waves. In addition to the expected resonances, such as the $a_4(1970)$ and $\pi(1800)$, we observe further signals in the $J^{PC}=3^{++}$ and $J^{PC}=6^{++}$ sectors, where some candidates are listed in the PDG, but no state is yet established. As predicted by lattice-QCD, we observe signals for $b_1(1235)\pi$ waves in the $J^{PC}=1^{-+}$ sector consistent with the $\pi_1(1600)$. We also observe a signal in $\rho\omega$ in the $J^{PC}=1^{-+}$ sector. To validate these promising signals and to quantify their parameters, the next step of this analysis is to model the mass dependence of the extracted partial-wave amplitudes.

In addition to the analysis of the $\omega\pi^-\pi^0$ final state, we investigate further channels. These include the $K_SK^-$ channel, which is the only channel in which an $a_6$ state has been observed so far. The $K_SK_S\pi$ channel is a good candidate for investigating the nature of the $a_1(1420)$. It also gives access to the $K^*\bar{K}$ channel, predicted to have a significant partial decay width of the $\pi_1(1600)$. The $f_1(1285)\pi$ channel could give access to the $\pi_1(1600)$ as it is predicted to have the second largest partial decay width. Its decay to $\eta\pi\pi\pi$ is being studied. By including the latter two decay channels, we plan to study the $\pi_1(1600)$ in practically all of its decay channels at COMPASS and thus obtain a complete picture of this state. This is only possible because of the unique COMPASS data set.

\acknowledgments
The research was funded by the DFG under Germany’s Excellence Strategy - EXC2094 - 390783311 and BMBF Verbundforschung 05P21WOCC1 COMPASS.


\begin{thebibliography}{0}
\bibitem{Dudek:2013yja} \BY{Dudek J. J. \etal} \IN{Phys. Rev. D}{88}{no.9}{2013}{094505},\\ DOI: 10.1103/PhysRevD.88.094505.
\bibitem{Woss:2020ayi} \BY{Woss A. J. \etal} \IN{Phys. Rev. D}{103}{no.5}{2021}{054502},\\ DOI: 10.1103/PhysRevD.103.054502.
\bibitem{Rodas:2018owy} \BY{Rodas A. \etal} \IN{Phys. Rev. Lett.}{122}{2019}{042002},\\ DOI: 10.1103/PhysRevLett.122.042002.
\bibitem{Akhunzyanov:2018lqa} \BY{Aghasyan M. \etal} \IN{Phys. Rev. D}{98}{2018}{092003},\\ DOI: 10.1103/PhysRevD.98.092003.
\bibitem{Workman:2022ynf} \BY{Workman R. L. \etal} \IN{PTEP}{2022}{2022}{083C01},\\ DOI: 10.1093/ptep/ptac097.
\bibitem{Cleland:1982te} \BY{Cleland W. E. \etal} \IN{Nucl. Phys. B}{208}{1982}{228-261},\\ DOI: 10.1016/0550-3213(82)90114-6.


\end{thebibliography}
\end{document}